\documentstyle[amssymb,floats,aps,prb,epsf]{revtex}
\epsfclipon
\begin{document}
\twocolumn[\hsize\textwidth\columnwidth\hsize\csname
@twocolumnfalse\endcsname

\draft
\title{Thermal conductivity in the doped two-leg ladder antiferromagnet
Sr$_{14-x}$Ca$_{x}$Cu$_{24}$O$_{41}$}

\author{Jihong Qin and Shiping Feng}
\address{Department of Physics, Beijing Normal University, Beijing 100875,
China}
\author{Feng Yuan}
\address{Texas Center for Superconductivity, University of Houston,
Houston, Texas 77204, U.S.A.}
\author{Wei Yeu Chen}
\address{Department of Physics, Tamkang University, Tamsui 25137,
Taiwan}
\maketitle
\begin{abstract}
Within the t-J model, the heat transport of the doped two-leg
ladder material Sr$_{14-x}$Ca$_{x}$Cu$_{24}$O$_{41}$ is studied in
the doped regime where superconductivity appears under pressure in
low temperatures. The energy dependence of the thermal
conductivity $\kappa_{c}(\omega)$ shows a low-energy peak, while
the temperature dependence of the thermal conductivity
$\kappa_{c}(T)$ is characterized by a broad band. In particular,
$\kappa_{c}(T)$ increases monotonously with increasing temperature
at low temperatures $T<0.05J$, and is weakly temperature dependent
in the temperature range $0.05J<T<0.1J$, then decreases for
temperatures $T>0.1J$, in qualitative agreement with experiments.
Our result also shows that although both dressed holons and
spinons are responsible for the thermal conductivity, the
contribution from the dressed spinons dominates the heat transport
of Sr$_{14-x}$Ca$_{x}$Cu$_{24}$O$_{41}$.
\end{abstract}
\pacs{74.25.Fy,74.62.Dh,74.72.Jt}

]
\bigskip
\narrowtext

In recent years the two-leg ladder material
Sr$_{14}$Cu$_{24}$O$_{41}$ has attracted great interest since its
ground state may be a spin liquid state with a finite spin gap
\cite{dagotto}. This spin liquid state may play a crucial role in
superconductivity of doped cuprates as emphasized by Anderson
\cite{anderson1}. When carriers are doped into
Sr$_{14}$Cu$_{24}$O$_{41}$, such as the isovalent substitution of
Ca for Sr, a metal-insulator transition occurs
\cite{dagotto,uehara,nagata}, and further, this doped two-leg
ladder material Sr$_{14-x}$Ca$_{x}$Cu$_{24}$O$_{41}$ is a
superconductor under pressure in low temperatures
\cite{uehara,nagata}. Apart from the observation of
superconductivity under pressure in
Sr$_{14-x}$Ca$_{x}$Cu$_{24}$O$_{41}$, the particular geometrical
arrangement of the Cu ions provides a playground for magnetic and
transport studies of low-dimensional strongly correlated materials
\cite{dagotto}. All cuprate superconductors found up now contain
square CuO$_{2}$ planes \cite{kastner}, whereas
Sr$_{14-x}$Ca$_{x}$Cu$_{24}$O$_{41}$ consists of two-leg ladders
of other Cu ions and edge-sharing CuO$_{2}$ chains
\cite{dagotto,uehara,nagata}, and is the only known
superconducting copper oxide without a square lattice.
Experimentally, it has been shown by virtue of the nuclear
magnetic resonance and nuclear quadrupole resonance, particularly
inelastic neutron scattering measurements that there is a region
of parameter space and doping where
Sr$_{14-x}$Ca$_{x}$Cu$_{24}$O$_{41}$ in the normal state is an
antiferromagnet with commensurate short-range order
\cite{dagotto,katano,magishi}. Moreover, transport measurements on
Sr$_{14-x}$Ca$_{x}$Cu$_{24}$O$_{41}$ in the same region of
parameter space and doping indicate that the resistivity is linear
with temperatures \cite{nagata}, one of the hallmarks of the
exotic normal state properties found in doped cuprates on a square
lattice \cite{kastner}. These unusual normal state properties do
not fit in the conventional Fermi-liquid theory \cite{anderson2},
and may be interpreted within the framework of the charge-spin
separation \cite{anderson1}, where the electron is separated into
a neutral spinon and a charged holon, then the basic excitations
of the system are not fermionic quasiparticles as in other
conventional metals with charge, spin and heat all carried by one
and the same particles.

The heat transport is one of the basic transport properties that
provide a wealth of useful information on carriers and phonons as
well as their scattering processes
\cite{uher,sologubenko,hess,kudo}. In the conventional metals, the
thermal conductivity contains both contributions from carriers and
phonons \cite{uher}. The phonon contribution to the thermal
conductivity is always present in the conventional metals, while
the magnitude of the carrier contribution depends on the type of
material because it is directly proportional to the free carrier
density. For the underdoped cuprates on a square lattice, the
phonon contribution to the thermal conductivity is strongly
suppressed \cite{nakamura,baberski}. Recently, an unusual
contribution to the thermal conductivity of the doped two-leg
ladder material Sr$_{14-x}$Ca$_{x}$Cu$_{24}$O$_{41}$ has been
observed \cite{sologubenko,hess,kudo}. It has been argued that
this unusual contribution may be due to an energy transport via
magnetic excitations \cite{sologubenko,gros}, and therefore can
not be explained within the conventional models of phonon heat
transport based on phonon-defect scattering or conventional
phonon-electron scattering. In particular, in the doped regime
where superconductivity appears under pressure in low
temperatures, this unusual contribution dominates the thermal
conductivity of Sr$_{14-x}$Ca$_{x}$Cu$_{24}$O$_{41}$. This is a
challenge issue since it is closely related to the doped Mott
insulating state that forms the basis for superconductivity
\cite{dagotto}. We \cite{feng1} have developed a charge-spin
separation fermion-spin theory to study the physical properties of
doped Mott insulators, where the electron operator is decoupled as
the gauge invariant dressed holon and spinon. Within this theory,
we \cite{feng2,feng3} have discussed charge transport and spin
response of Sr$_{14-x}$Ca$_{x}$Cu$_{24}$O$_{41}$. It has been
shown that the charge transport is mainly governed by the
scattering from the dressed holons due to the dressed spinon
fluctuation, while the scattering from the dressed spinons due to
the dressed holon fluctuation dominates the spin response
\cite{feng1}. In this paper, we apply this successful approach to
discuss the heat transport of
Sr$_{14-x}$Ca$_{x}$Cu$_{24}$O$_{41}$. Within the $t$-$J$ model, we
show that although both dressed holons and spinons are responsible
for the heat transport, the contribution from the dressed spinons
dominates the thermal conductivity of
Sr$_{14-x}$Ca$_{x}$Cu$_{24}$O$_{41}$ in the doped regime where
superconductivity appears under pressure in low temperatures.

The two-leg ladder is defined as two parallel chains of ions, with
bonds among them so that the interchain coupling is comparable in
strength to the couplings along the chains, while the coupling
between the two chains that participates in this structure is
through rungs \cite{dagotto}. It has been argued that the
essential physics of the doped two-leg ladder antiferromagnet is
contained in the two-leg ladder, and can be effectively described
by the $t$-$J$ model \cite{dagotto},
\begin{eqnarray}
H&=&-t_{\parallel}\sum_{i\hat{\eta}a\sigma}C_{ia\sigma}^{\dagger}
C_{i+\hat{\eta}a\sigma}-t_{\perp}\sum_{i\sigma}
(C_{i1\sigma}^{\dagger}C_{i2\sigma}+C_{i2\sigma}^{\dagger}
C_{i1\sigma})\nonumber \\
&-&\mu\sum_{ia\sigma}C_{ia\sigma}^{\dagger} C_{ia\sigma}\nonumber \\
&+&J_{\parallel}\sum_{i\hat{\eta}a}{\bf S}_{ia}\cdot {\bf
S}_{i+\hat{\eta}a}+J_{\perp}\sum_{i}{\bf S}_{i1}\cdot {\bf
S}_{i2},
\end{eqnarray}
supplemented by a single occupancy local constraint $\sum_{\sigma}
C_{ia\sigma}^{\dagger}C_{ia\sigma}\leq 1$, where $\hat{\eta}=\pm
c_{0}\hat{x}$, $c_{0}$ is the lattice constant of the two-leg
ladder lattice, which is set as unity hereafter, $i$ runs over all
rungs, $\sigma(=\uparrow,\downarrow)$ and $a(=1,2)$ are spin and
leg indices, respectively, $C^{\dagger}_{ia\sigma}$
($C_{ia\sigma}$) is the electron creation (annihilation) operator,
${\bf S}_{ia}=C^{\dagger}_{ia} \vec{\sigma}C_{ia}/2$ is the spin
operator with ${\vec\sigma}=(\sigma_{x},\sigma_{y},\sigma_{z})$ as
the Pauli matrices, and $\mu$ is the chemical potential. In the
materials of interest \cite{uehara,katano}, the exchange coupling
$J_{\parallel}$ along the legs is close to the exchange coupling
$J_{\perp}$ across a rung, and the same is true of the hopping
$t_{\parallel}$ along the legs and the rung hopping strength
$t_{\perp}$. Therefore in the following discussions, we will work
with the isotropic system $J_{\perp}=J_{\parallel}=J$, $t_{\perp}=
t_{\parallel}=t$. The strong electron correlation in the $t$-$J$
model manifests itself by the single occupancy local constraint,
and thus the crucial requirement is to impose this local
constraint. This local constraint can be treated properly within
the charge-spin separation fermion-spin theory \cite{feng1},
$C_{ia\uparrow}= h^{\dagger}_{ia\uparrow}S^{-}_{ia}$,
$C_{ia\downarrow}= h^{\dagger}_{ia\downarrow}S^{+}_{ia}$, where
the spinful fermion operator $h_{ia\sigma}=e^{-i\Phi_{i\sigma}}
h_{ia}$ describes the charge degree of freedom together with some
effects of the spinon configuration rearrangements due to the
presence of the hole itself (dressed holon), while the spin
operator $S_{ia}$ describes the spin degree of freedom (dressed
spinon), then the electron local constraint for the single
occupancy, $\sum_{\sigma} C^{\dagger}_{ia\sigma} C_{ia\sigma}=
S^{+}_{ia}h_{ia\uparrow}h^{\dagger}_{ia\uparrow}S^{-}_{ia} +
S^{-}_{ia}h_{ia\downarrow}h^{\dagger}_{ia\downarrow}S^{+}_{ia}
=h_{ia}h^{\dagger}_{ia}(S^{+}_{ia} S^{-}_{ia}+S^{-}_{ia}
S^{+}_{ia})=1- h^{\dagger}_{ia}h_{ia}\leq 1$, is satisfied in
analytical calculations, and the double spinful fermion occupancy,
$h^{\dagger}_{ia\sigma}h^{\dagger}_{ia-\sigma}
=e^{i\Phi_{i\sigma}} h^{\dagger}_{ia}h^{\dagger}_{ia}
e^{i\Phi_{i-\sigma}}=0$, $h_{ia\sigma} h_{ia-\sigma}=
e^{-i\Phi_{i\sigma}}h_{ia}h_{ia}e^{-i\Phi_{i-\sigma}}=0$, are
ruled out automatically. It has been emphasized that this dressed
holon $h_{ia\sigma}$ is a spinless fermion $h_{ia}$ incorporated a
spinon cloud $e^{-i\Phi_{i\sigma}}$ (magnetic flux), and then is a
magnetic dressing. In other words, the gauge invariant dressed
holon carries some spinon messages, i.e., it shares its nontrivial
spinon environment \cite{martins}. Although in common sense
$h_{ia\sigma}$ is not a real spinful fermion, it behaves like a
spinful fermion. In this charge-spin separation fermion-spin
representation, the low-energy behavior of the $t$-$J$ ladder (1)
can be expressed as \cite{feng1},
\begin{eqnarray}
H&=&t\sum_{i\hat{\eta}a}(h^{\dagger}_{i+\hat{\eta}a\uparrow}
h_{ia\uparrow}S^{+}_{ia}S^{-}_{i+\hat{\eta}a}+
h^{\dagger}_{i+\hat{\eta}a\downarrow}h_{ia\downarrow}S^{-}_{ia}
S^{+}_{i+\hat{\eta}a})\nonumber \\
&+&t\sum_{i}(h^{\dagger}_{i2\uparrow}h_{i1\uparrow}S^{+}_{i1}
S^{-}_{i2}+h^{\dagger}_{i2\downarrow}h_{i1\downarrow}S^{-}_{i1}
S^{+}_{i2}) \nonumber \\
&+&\mu\sum_{ia\sigma}h^{\dagger}_{ia\sigma}h_{ia\sigma}\nonumber \\
 &+&J_{{\rm eff}}\sum_{i\hat{\eta}a}{\bf S}_{ia}\cdot {\bf S}_{i+\hat{\eta}a}+
J_{{\rm eff}}\sum_{i}{\bf S}_{i1} \cdot {\bf S}_{i2},
\end{eqnarray}
with $J_{\rm eff}=J(1-\delta)^{2}$, and $\delta=\langle
h^{\dagger}_{ia\sigma}h_{ia\sigma}\rangle=\langle h^{\dagger}_{ia}
h_{ia}\rangle$ is the hole doping concentration. In this case, the
kinetic part has been expressed as the dressed holon-spinon
interaction, and therefore reflect a competition between the
kinetic energy and magnetic energy. This competition dominates the
essential physics since the dressed holon and spinon self-energies
are ascribed purely to the dressed holon-spinon interaction.

Now we follow the linear response theory \cite{mahan}, and obtain
the thermal conductivity along the ladder,
\begin{eqnarray}
\kappa_{c}(\omega,T)=-{1\over T}{{\rm Im}\Pi_{Q} (\omega,T)\over
\omega},
\end{eqnarray}
with $\Pi_{Q}(\omega,T)$ is the heat current-current correlation
function, and is defined as,
\begin{eqnarray}
\Pi_{Q}(\tau-\tau')=-\langle T_{\tau}j_{Q}(\tau)j_{Q}(\tau')
\rangle,
\end{eqnarray}
where $\tau$ and $\tau'$ are the imaginary times and $T_{\tau}$ is
the $\tau$ order operator, while the heat current density is
obtained within the $t$-$J$ ladder Hamiltonian (2) by using
Heisenberg's equation of motion as,
\begin{mathletters}
\begin{eqnarray}
j_{Q}&=&i\sum_{i,j}\sum_{a,b}{\bf
R}_{jb}[H_{ia},H_{jb}]=j^{(h)}_{Q}+
j^{(s)}_{Q}, \\
j^{(h)}_{Q}&=&-i(\chi_{\parallel} t)^{2}
\sum_{i\hat{\eta}\hat{\eta}'a\sigma}\hat{\eta}h_{i+\hat{\eta}+
\hat{\eta}'a\sigma}^{\dagger}h_{ia\sigma} \nonumber \\
&+&i\chi_{\parallel}\chi_{\perp}t^{2}\sum_{i\hat{\eta}\sigma}
[({\bf R}_{i2}-{\bf R}_{i1}-\hat{\eta})
h_{i+\hat{\eta}1\sigma}^{\dagger}h_{i2\sigma}\nonumber\\
&-& ({\bf R}_{i2}-{\bf R}_{i1}+\hat{\eta})
h_{i+\hat{\eta}2\sigma}^{\dagger}h_{i1\sigma}] \nonumber\\
&-&i\mu\chi_{\parallel} t\sum_{i\hat{\eta}a\sigma}\hat{\eta}
h_{i+\hat{\eta}a\sigma}^{\dagger}h_{ia\sigma}\nonumber\\
&+&i\mu\chi_{\perp}t\sum_{i\sigma} ({\bf R}_{i2}-{\bf
R}_{i1})(h_{i1\sigma}^{\dagger}h_{i2\sigma}
-h_{i2\sigma}^{\dagger}h_{i1\sigma}) \nonumber\\
&+&i(\chi_{\perp}t)^{2}\sum_{i\sigma}({\bf R}_{i2}- {\bf
R}_{i1})(h_{i2\sigma}^{\dagger}h_{i2\sigma}-
h_{i1\sigma}^{\dagger}h_{i1\sigma}),\\
j^{(s)}_{Q}&=&i{1\over 2}\epsilon_{\parallel}J_{\rm{eff}}^{2}
\sum_{i\hat{\eta}\hat{\eta}'a}(\hat{\eta}'-\hat{\eta})\nonumber\\
&[&\epsilon_{\parallel} S^{z}_{ia}(S^{+}_{i+\hat{\eta}'a}
S^{-}_{i+\hat{\eta}a}- S^{-}_{i+\hat{\eta}'a}
S^{+}_{i+\hat{\eta}a}) \nonumber \\
&-& S^{z}_ {i+\hat{\eta}'a}(S^{+}_{ia}S^{-}_{i+\hat{\eta}a}-
S^{-}_{ia}S^{+}_{i+\hat{\eta}a}) \nonumber\\
&+&(S^{+}_{ia}S^{-}_{i+\hat{\eta}'a}-S^{-}_{ia}S^{+}_{i+\hat{\eta}'a})
S^{z}_ {i+\hat{\eta}a}]\nonumber \\
&+&i{1\over 2}J_{\rm{eff}}^{2}\sum_{i\hat{\eta}}
\{\hat{\eta}\epsilon_{\parallel}\epsilon_{\perp}[S^{z}_{i1}
(S^{+}_{i+\hat{\eta}1}S^{-}_{i2}-S^{+}_{i2}
S^{-}_{i+\hat{\eta}1})\nonumber\\
&+&S^{z}_{i2}(S^{+}_{i+\hat{\eta}2}
S^{-}_{i1}-S^{+}_{i1}S^{-}_{i+\hat{\eta}2})] \nonumber \\
&+&\hat{\eta}\epsilon_{\parallel}[S^{z}_{i1} (S^{+}_{i2}
S^{-}_{i+\hat{\eta}2}-S^{-}_{i2}S^{+}_{i+\hat{\eta}2})\nonumber\\
&+&S^{z}_{i2}(S^{+}_{i1}S^{-}_{i+\hat{\eta}1}-S^{-}_{i1}
S^{+}_{i+\hat{\eta}1})] \nonumber\\
&+&\hat{\eta}\epsilon_{\perp}[(S^{z}_{i+\hat{\eta}1}-
S^{z}_{i+\hat{\eta}2})(S^{+}_{i2}S^{-}_{i1}-S^{+}_{i1}
S^{-}_{i2})]\nonumber \\
&+&({\bf R}_{i2}-{\bf R}_{i1})\epsilon_{\parallel}[S^{z}_{i1}
(S^{+}_{i2}S^{-}_{i+\hat{\eta}2}-S^{-}_{i2}
S^{+}_{i+\hat{\eta}2})\nonumber\\
&-&S^{z}_{i2}(S^{+}_{i1}
S^{-}_{i+\hat{\eta}1}-S^{-}_{i1}S^{+}_{i+\hat{\eta}1})]
\nonumber \\
&+&2({\bf R}_{i2}-{\bf R}_{i1})\epsilon_{\parallel}
\epsilon_{\perp}[S^{z}_{i1}S^{+}_{i2}S^{-}_{i+\hat{\eta}1}-
S^{z}_{i2}S^{+}_{i1}S^{-}_{i+\hat{\eta}2}] \nonumber \\
&+&2({\bf R}_{i2}-{\bf R}_{i1})\epsilon_{\perp}
[S^{z}_{i+\hat{\eta}2}S^{+}_{i1}S^{-}_{i2}-
S^{z}_{i+\hat{\eta}1}S^{+}_{i2}S^{-}_{i1}]\} \nonumber \\
&+&i{1\over 4} {\epsilon_{\perp}}^{2}J_{\rm{eff}}^{2}\sum_{i}
({\bf R}_{i2}-{\bf R}_{i1})(S^{+}_{i1}S^{-}_{i1}-S^{+}_{i2}
S^{-}_{i2} ),
\end{eqnarray}
\end{mathletters}
where ${\bf R}_{i1}$ and ${\bf R}_{i2}$ are lattice sites of leg 1
and leg 2, respectively,
$\epsilon_{\parallel}=1+2t\phi_{\parallel}/J_{\rm eff}$,
$\epsilon_{\perp}=1+4t\phi_{\perp}/J_{\rm eff}$, the dressed
spinon correlation functions $\chi_{\parallel}=\langle S_{ai}^{+}
S_{ai+\hat{\eta}}^{-}\rangle$, $\chi_{\perp}=\langle S^{+}_{1i}
S^{-}_{2i}\rangle$, and the dressed holon particle-hole order
parameters $\phi_{\parallel}=\langle h^{\dagger}_{ai\sigma}
h_{ai+\hat{\eta}\sigma}\rangle$, $\phi_{\perp}=\langle
h^{\dagger}_{1i\sigma}h_{2i\sigma}\rangle$. Although the total
heat current density $j_{Q}$ has been separated into two parts
$j^{(h)}_{Q}$ and $j^{(s)}_{Q}$, with $j^{(h)}_{Q}$ is the dressed
holon heat current density, and $j^{(s)}_{Q}$ is the dressed
spinon heat current density, the strong correlation between
dressed holons and spinons is still included self-consistently
through the dressed spinon's order parameters entering in the
dressed holon's propagator, and the dressed holon's order
parameters entering in the dressed spinon's propagator. In this
case, the heat current-current correlation function of the two-leg
ladder system can be obtained in terms of the full dressed holon
and spinon Green's functions $g_{\sigma}(k,\omega)$ and
$D(k,\omega)$. Because there are two coupled $t$-$J$ chains in the
two-leg ladder system, the energy spectrum has two branches.
Therefore the one-particle dressed holon and spinon Green's
functions are matrices and can be expressed as
$g_{\sigma}(i-j,\tau-\tau')=g_{L\sigma}(i-j,\tau-\tau')+
\sigma_{x}g_{T\sigma}(i-j,\tau-\tau')$ and
$D(i-j,\tau-\tau^{\prime})=D_{L}(i-j,\tau-\tau^{\prime})+\sigma_{x}
D_{T}(i-j,\tau-\tau^{\prime})$, respectively, where the
longitudinal and transverse parts are defined as
$g_{L\sigma}(i-j,\tau-\tau')=-\langle T_{\tau}h_{ia\sigma}(\tau)
h^{\dagger}_{ja\sigma}(\tau')\rangle$,
$D_{L}(i-j,\tau-\tau^{\prime})= -\langle T_{\tau}S_{ia}^{+}
(\tau)S_{ja}^{-}(\tau^{\prime})\rangle$ and
$g_{T\sigma}(i-j,\tau-\tau')=-\langle T_{\tau}h_{ia\sigma}
(\tau)h^{\dagger}_{ja'\sigma}(\tau')\rangle$, $D_{T}
(i-j,\tau-\tau^{\prime})=-\langle T_{\tau}S_{ia}^{+}(\tau)
S_{ja^{\prime}}^{-}(\tau^{\prime})\rangle$, with $a'\neq a$.
Following the discussions of the charge transport \cite{feng2}, we
can obtain the thermal conductivity of the doped two-leg ladder
antiferromagnet as,
\begin{mathletters}
\begin{eqnarray}
\kappa_{c}(\omega,T)&=&\kappa^{(h)}_{c}(\omega,T)+
\kappa^{(s)}_{c}(\omega,T), \\
\kappa^{(h)}_{c}(\omega,T)&=&-{1\over N}\sum_{k\sigma}
\int^{\infty}_{-\infty}{d\omega'\over 2\pi} \nonumber \\
&\times& \{\Lambda_{h1} [A_{T}^{(h\sigma)}(k,\omega'+\omega)
A_{T}^{(h\sigma)}(k,\omega')\nonumber\\
&+&A_{L}^{(h\sigma)}(k,\omega'+\omega)
A_{L}^{(h\sigma)}(k,\omega')]\nonumber \\
&+&\Lambda_{h2} [A_{T}^{(h\sigma)}(k,\omega'+\omega)
A_{L}^{(h\sigma)}(k,\omega')\nonumber\\
&+&A_{L}^{(h\sigma)}(k,\omega'+\omega)
A_{T}^{(h\sigma)}(k,\omega')]\nonumber\\
&+&\Lambda_{h3} [A_{T}^{(h\sigma)}(k,\omega'+\omega)
A_{T}^{(h\sigma)}(k,\omega')\nonumber\\
&-&A_{L}^{(h\sigma)}(k,\omega'+\omega)
A_{L}^{(h\sigma)}(k,\omega')]\}\nonumber\\
&\times &{n_{F}(\omega'+\omega)-n_{F}(\omega') \over T\omega},\\
\kappa^{(s)}_{c}(\omega,T)&=&-{1\over 2N}\sum_{k}
\int^{\infty}_{-\infty}{d\omega'\over 2\pi}\nonumber\\
&\times&\{\Lambda_{s1} [A_{L}^{(s)}(k,\omega'+\omega)
A_{L}^{(s)}(k,\omega')\nonumber\\
&+&A_{T}^{(s)}(k,\omega'+\omega) A_{T}^{(s)}
(k,\omega')]\nonumber \\
&+&\Lambda_{s2} [A_{L}^{(s)}(k,\omega'+\omega) A_{T}^{(s)}
(k,\omega')\nonumber\\
&+&A_{T}^{(s)}(k,\omega'+\omega) A_{L}^{(s)}
(k,\omega')]\nonumber \\
&+&\Lambda_{s3} [A_{L}^{(s)}(k,\omega'+\omega) A_{L}^{(s)}
(k,\omega')\nonumber\\
&-&A_{T}^{(s)}(k,\omega'+\omega)A_{T}^{(s)}(k,\omega')]\}
\nonumber \\
&\times &{n_{B}(\omega'+\omega)-n_{B}(\omega') \over T\omega},
\end{eqnarray}
\end{mathletters}
where $\kappa^{(h)}_{c}(\omega,T)$ and $\kappa^{(s)}_{c}
(\omega,T)$ are the corresponding contributions from dressed
holons and spinons, respectively, $N$ is the number of rungs, and
\begin{mathletters}
\begin{eqnarray}
\Lambda_{h1}&=&{(Z\chi_{\parallel} t)}^2\gamma^{2}_{sk}
[(Z\chi_{\parallel} t\gamma_{k}+\mu)^2+ (\chi_{\perp} t)^2], \\
\Lambda_{h2}&=&2{(Z\chi_{\parallel} t)}^2\gamma^{2}_{sk}
(Z\chi_{\parallel} t\gamma_{k}+\mu)(\chi_{\perp} t), \\
\Lambda_{h3}&=&-(\chi_{\perp} t)^2[(Z\chi_{\parallel}
t\gamma_{k}+\mu)^2- (\chi_{\perp} t)^2], \\
\Lambda_{s1}&=&2\gamma^{2}_{sk}J_{\rm eff}^4Z^2
\{[2Z\epsilon_{\parallel} (\epsilon_{\parallel}
\chi_{\parallel}+C_{\parallel} -2\chi_{\parallel}
\gamma_{k})\nonumber\\
&+&\epsilon_{\perp} (\epsilon_{\parallel}
\chi_{\perp}+C_{\perp})]^2
+(\epsilon_{\parallel}\chi_{\perp}+\epsilon_{\perp}
\chi_{\parallel})^2\}, \\
\Lambda_{s2}&=&-4\gamma^{2}_{sk}J_{\rm eff}^4Z^2\nonumber\\
&\{&2Z \epsilon_{\parallel}(\epsilon_{\parallel}\chi_{\perp}+
\epsilon_{\perp}\chi_{\parallel}) (\epsilon_{\parallel}
\chi_{\parallel}+C_{\parallel} -2\chi_{\parallel}
\gamma_{k})\nonumber \\
&+& \epsilon_{\perp} (\epsilon_{\parallel} \chi_{\perp}
+C_{\perp})(\epsilon_{\parallel}\chi_{\perp}+
\epsilon_{\perp}\chi_{\parallel})\}, \\
\Lambda_{s3}&=&-2J_{\rm eff}^4\{[{1\over 4}
\epsilon_{\perp}^2\nonumber\\
&+&Z\epsilon_{\perp}((\epsilon_{\parallel}
\chi_{\perp}-C_{\perp})\gamma_{k}+(\epsilon_{\parallel}
C_{\perp}-\chi_{\perp}))]^2 \nonumber \\
&-&Z ^2[(\epsilon_{\parallel}\chi_{\perp}+\epsilon_{\perp}
\chi_{\parallel})\gamma_{k} -\epsilon_{\parallel}
(\epsilon_{\perp}\chi_{\parallel}+ C_{\perp})]^2 \},
\end{eqnarray}
\end{mathletters}
with $\gamma^{2}_{sk}={\rm sin}^{2}k_{x}$, $\gamma_{k}={\rm
cos}k_{x}$, $Z$ is the coordination number within the leg, the
dressed spinon correlation functions $C_{\parallel}=(1/Z^{2})
\sum_{\hat{\eta}\hat{\eta'}}\langle S_{ai+\hat{\eta}}^{+}
S_{ai+\hat{\eta'}}^{-}\rangle$, $C_{\perp}=(1/Z)\sum_{\hat{\eta}}
\langle S_{2i}^{+}S_{1i+\hat{\eta}}^{-}\rangle$, $n_{F}(\omega)$
and $n_{B}(\omega)$ are the fermion and boson distribution
functions, respectively, and the longitudinal and transverse
spectral functions of the dressed holon and spinon are obtained as
$A_{L}^{(h\sigma)}(k,\omega)=-2 {\rm Im}g_{L\sigma}(k,\omega)$,
$A_{L}^{(s)}(k,\omega)=-2 {\rm Im}D_{L}(k,\omega)$ and
$A_{T}^{(h\sigma)}(k,\omega)=-2 {\rm Im}g_{T\sigma}(k,\omega)$,
$A_{T}^{(s)}(k,\omega)=-2 {\rm Im}D_{T}(k,\omega)$, respectively,
where the full dressed holon and spinon Green's functions have
been obtained in Refs. \cite{feng2,feng3}, and are expressed as,
$g^{-1}_{\sigma}(k,\omega)=g^{(0)-1}_{\sigma}(k,\omega)-
\Sigma^{(h)}(k,\omega)$ and $D^{-1}(k,\omega)= D^{(0)-1}
(k,\omega)-\Sigma^{(s)}(k,\omega)$, with the longitudinal and
transverse mean-field dressed holon and spinon Green's functions
$g^{(0)}_{L\sigma}(k,\omega)=1/2\sum_{\nu}1/(\omega-
\xi^{(\nu)}_{k})$, $D^{(0)}_{L}(k,\omega)={1/2}\sum_{\nu}
{B^{(\nu)}_{k} / (\omega^{2}-(\omega^{(\nu)}_{k})^{2})}$ and
$g^{(0)}_{T\sigma}(k,\omega)=1/2\sum_{\nu} (-1)^{\nu+1}
/(\omega-\xi^{(\nu)}_{k})$, $D^{(0)}_{T}(k,\omega)={1/2}
\sum_{\nu}(-1)^{\nu+1} {B^{(\nu)}_{k} /(\omega^{2}-
(\omega^{(\nu)}_{k})^{2})}$, where $\nu=1,2$, and the longitudinal
and transverse second-order dressed holon and spinon self-energies
are obtained by the loop expansion to the
second-order\cite{feng2,feng3} as,
\begin{mathletters}
\begin{eqnarray}
\Sigma_{L}^{(h)}(k,\omega)=&(&{t\over N})^{2}\sum_{pq}
\sum_{\nu\nu'\nu''}\Xi^{(h)}_{\nu\nu'\nu''}(k,p,q,\omega), \\
\Sigma_{T}^{(h)}(k,\omega)=&(&{t\over N})^{2}\sum_{pq}
\sum_{\nu\nu'\nu''}\nonumber\\
&(&-1)^{\nu+\nu'+\nu''+1}
\Xi^{(h)}_{\nu\nu'\nu''}(k,p,q,\omega),\\
\Sigma_{L}^{(s)}(k,\omega)=&(&{t\over N})^{2}\sum_{pq}
\sum_{\nu\nu'\nu''}\Xi^{(s)}_{\nu\nu'\nu''}(k,p,q,\omega), \\
\Sigma_{T}^{(s)}(k,\omega)=&(&{t\over N})^{2}\sum_{pq}
\sum_{\nu\nu'\nu''}\nonumber\\
&(&-1)^{\nu+\nu'+\nu''+1} \Xi^{(s)}_{\nu\nu'\nu''}(k,p,q,\omega),
\end{eqnarray}
\end{mathletters}
respectively, where $\Xi^{(h)}_{\nu\nu'\nu''}(k,p,q,\omega)$ and
$\Xi^{(s)}_{\nu\nu'\nu''}(k,p,q,\omega)$ are given by,
\begin{mathletters}
\begin{eqnarray}
\Xi^{(h)}_{\nu\nu'\nu''}(k,p,q,\omega)&=&{B^{(\nu')}_{q+p}
B^{(\nu)}_{q}\over 32\omega^{(\nu')}_{q+p}\omega^{(\nu)}_{q}}{1
\over 2} \{[Z\gamma_{q+p+k}+(-1)^{\nu+\nu''}]^{2}\nonumber\\
&+&[Z\gamma_{q-k}+(-1)^{\nu'+\nu''}]^{2}\} \nonumber \\
&\times&({F^{(1)}_{\nu\nu'\nu''}(k,p,q)\over\omega+
\omega^{(\nu')}_{q+p}-\omega^{(\nu)}_{q}-\xi^{(\nu'')}_{p+k}}\nonumber\\
&+&{F^{(2)}_{\nu\nu'\nu''}(k,p,q)\over\omega-
\omega^{(\nu')}_{q+p}+\omega^{(\nu)}_{q}-\xi^{(\nu'')}_{p+k}}
\nonumber \\
&+&{F^{(3)}_{\nu\nu'\nu''}(k,p,q)\over\omega+
\omega^{(\nu')}_{q+p}+\omega^{(\nu)}_{q}-\xi^{(\nu'')}_{p+k}}\nonumber\\
&+&{F^{(4)}_{\nu\nu'\nu''}(k,p,q)\over\omega-
\omega^{(\nu')}_{q+p}-\omega^{(\nu)}_{q}-\xi^{(\nu'')}_{p+k}}),
\\
\Xi^{(s)}_{\nu\nu'\nu''}(k,p,q,\omega)&=&{B^{(\nu'')}_{k+p} \over
16\omega^{(\nu'')}_{k+p}} \{[Z\gamma_{q+p+k}+(-1)^{\nu+\nu''}]^{2}\nonumber\\
&+&[Z\gamma_{q-k}+(-1)^{\nu'+\nu''}]^{2}\} \nonumber \\
&\times&({\Gamma^{(1)}_{\nu\nu'\nu''}(k,p,q)
\over\omega+\xi^{(\nu')}_{p+q}-\xi^{(\nu)}_{q}-
\omega^{(\nu'')}_{k+p}}\nonumber\\
&-&{\Gamma^{(2)}_{\nu\nu'\nu''}(k,p,q)\over
\omega+\xi^{(\nu')}_{p+q} -\xi^{(\nu)}_{q}+
\omega^{(\nu'')}_{k+p}}),
\end{eqnarray}
\end{mathletters}
with $B^{(\nu)}_{k}=B_{k}-J_{\rm eff} [\chi_{\perp}+
2\chi^{z}_{\perp}(-1)^{\nu}][\epsilon_{\perp} +(-1)^{\nu}]$,
$B_{k}=\lambda [(2\epsilon_{\parallel} \chi^{z}_{\parallel}+
\chi_{\parallel}) \gamma_{k}- (\epsilon_{\parallel}
\chi_{\parallel}+ 2\chi^{z}_{\parallel})]$, $\lambda=2ZJ_{\rm
eff}$, and
\begin{mathletters}
\begin{eqnarray}
F^{(1)}_{\nu\nu'\nu''}&(&k,p,q)=n_{F}(\xi^{(\nu'')}_{p+k})[n_{B}
(\omega^{(\nu)}_{q})-n_{B}(\omega^{(\nu')}_{q+p})]\nonumber\\
&+&n_{B} (\omega^{(\nu')}_{q+p})[1+n_{B}(\omega^{(\nu)}_{q})]\\
F^{(2)}_{\nu\nu'\nu''}&(&k,p,q)=n_{F}(\xi^{(\nu'')}_{p+k})[n_{B}
(\omega^{(\nu')}_{q+p})-n_{B}(\omega^{(\nu)}_{q})]\nonumber\\
&+&n_{B} (\omega^{(\nu)}_{q})[1+n_{B}(\omega^{(\nu')}_{q+p})]\\
F^{(3)}_{\nu\nu'\nu''}&(&k,p,q)=n_{F}(\xi^{(\nu'')}_{p+k})[1+n_{B}
(\omega^{(\nu')}_{q+p})\nonumber\\
&+&n_{B}(\omega^{(\nu)}_{q})]
+n_{B} (\omega^{(\nu)}_{q})n_{B}(\omega^{(\nu')}_{q+p})\\
F^{(4)}_{\nu\nu'\nu''}&(&k,p,q)=[1+n_{B}(\omega^{(\nu)}_{q})]
[1+n_{B}(\omega^{(\nu')}_{q+p})]\nonumber\\
&-&n_{F}(\xi^{(\nu'')}_{p+k})[1+n_{B}
(\omega^{(\nu')}_{q+p})+n_{B}(\omega^{(\nu)}_{q})]\\
\Gamma^{(1)}_{\nu\nu'\nu''}&(&k,p,q) = n_{F}
(\xi^{(\nu')}_{p+q})[1-n_{F}(\xi^{(\nu)}_{q})]\nonumber\\
&-&n_{B} (\omega^{(\nu'')}_{k+p})[n_{F}(\xi^{(\nu)}_{q})-n_{F}
(\xi^{(\nu')}_{p+q})]\\
\Gamma^{(2)}_{\nu\nu'\nu''}&(&k,p,q)= n_{F}
(\xi^{(\nu')}_{p+q})[1-n_{F}(\xi^{(\nu)}_{q})]\nonumber\\
&+&[1+n_{B} (\omega^{(\nu'')}_{k+p})][n_{F}(\xi^{(\nu)}_{q})-n_{F}
(\xi^{(\nu')}_{p+q})],
\end{eqnarray}
\end{mathletters}
and the mean-field dressed holon and spinon excitations,
\begin{mathletters}
\begin{eqnarray}
\xi^{(\nu)}_{k}&=&Zt\chi_{\parallel}\gamma_{k}+ \mu+
\chi_{\perp}t(-1)^{\nu+1}, \\
\omega^{(\nu)2}_{k}&=&\alpha\epsilon_{\parallel}\lambda^{2}
({1\over 2}\chi_{\parallel}+\epsilon_{\parallel}
\chi^{z}_{\parallel})\gamma_{k}^{2}-\epsilon_{\parallel}
\lambda^{2}[{1\over 2}\alpha({1\over
2}\epsilon_{\parallel}\chi_{\parallel}\nonumber\\&+&
\chi^{z}_{\parallel}) +\alpha(C^{z}_{\parallel}+{1\over 2}
C_{\parallel})+{1\over 4}(1-\alpha)]\gamma_{k} \nonumber \\
&-&{1\over 2}\alpha\epsilon_{\perp}\lambda J_{{\rm eff}}
(C_{\perp}+\epsilon_{\parallel}\chi_{\perp})\gamma_{k}
+\alpha\lambda J_{{\rm eff}}(-1)^{\nu+1}\nonumber\\
&[&{1\over 2}(\epsilon_{\perp} \chi_{\parallel}
+\epsilon_{\parallel}\chi_{\perp})+
\epsilon_{\parallel}\epsilon_{\perp}(\chi^{z}_{\perp}+
\chi^{z}_{\parallel})]\gamma_{k} \nonumber \\
&-&\alpha\epsilon_{\parallel}\lambda J_{{\rm eff}}
(C^{z}_{\perp}+\chi^{z}_{\perp})\gamma_{k}+\lambda^{2}[\alpha
(C^{z}_{\parallel}+{1\over
2}\epsilon^{2}_{\parallel}C_{\parallel})\nonumber\\
&+&{1\over 8}(1-\alpha)(1+\epsilon^{2}_{\parallel}) -{1\over 2}
\alpha\epsilon_{\parallel}({1\over 2} \chi_{\parallel}+
\epsilon_{\parallel}\chi^{z}_{\parallel})]\nonumber \\
&+&\alpha\lambda J_{{\rm eff}}[\epsilon_{\parallel}
\epsilon_{\perp}C_{\perp}+2C^{z}_{\perp}]+{1\over 4} J^{2}_{{\rm
eff}}(\epsilon^{2}_{\perp}+1) \nonumber \\
&-&{1\over 2} \epsilon_{\perp}J^{2}_{{\rm eff}}(-1)^{\nu+1}
-\alpha\lambda J_{{\rm eff}}(-1)^{\nu+1}[{1\over 2}
\epsilon_{\parallel}\epsilon_{\perp}\chi_{\parallel}\nonumber
\\&+& \epsilon_{\perp}(\chi^{z}_{\parallel}+C^{z}_{\perp})+{1\over
2} \epsilon_{\parallel}C_{\perp}],
\end{eqnarray}
\end{mathletters}
where the dressed spinon correlation functions
$\chi^{z}_{\parallel}= \langle S_{ai}^{z}S_{ai+\hat{\eta}}^{z}
\rangle$, $\chi^{z}_{\perp}=\langle S_{1i}^{z}S_{2i}^{z}\rangle$,
$C^{z}_{\parallel}=(1/{Z^2})\sum_{\hat{\eta}\hat{\eta'}}\langle
S_{ai+\hat{\eta}}^{z} S_{ai+\hat{\eta'}}^{z}\rangle$, and
$C^{z}_{\perp}=(1/Z)\sum_{\hat{\eta}}\langle S_{1i}^{z}S_{2i+
\hat{\eta}}^{z}\rangle$. In order to satisfy the sum rule for the
correlation function $\langle S^{+}_{ai}S^{-}_{ai}\rangle =1/2$ in
the absence of the antiferromagnetic long-range-order, a
decoupling parameter $\alpha$ has been introduced in the
mean-field calculation, which can be regarded as the vertex
correction\cite{feng4}. All the above mean-field order parameters,
decoupling parameter $\alpha$, and chemical potential $\mu$, have
been determined by the self-consistent calculation.

\begin{figure}[prb]
\epsfxsize=3.0in\centerline{\epsffile{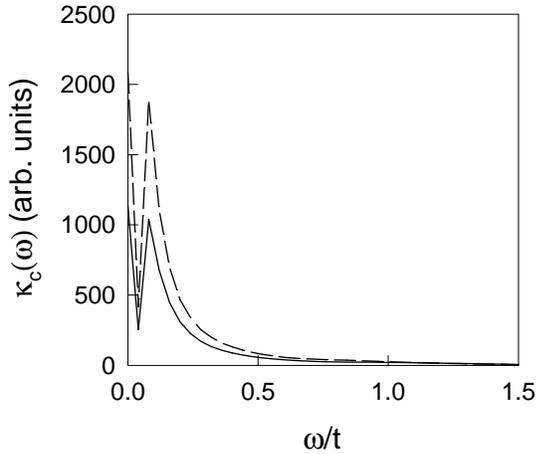}} \caption{The
thermal conductivity of the doped two-leg ladder material as a
function of frequency at doping $\delta=0.16$ (solid line) and
$\delta=0.20$ (dashed line) for parameter $t/J=2.5$ with
temperature $T=0.05J$.}
\end{figure}

\begin{figure}[prb]
\epsfxsize=3.0in\centerline{\epsffile{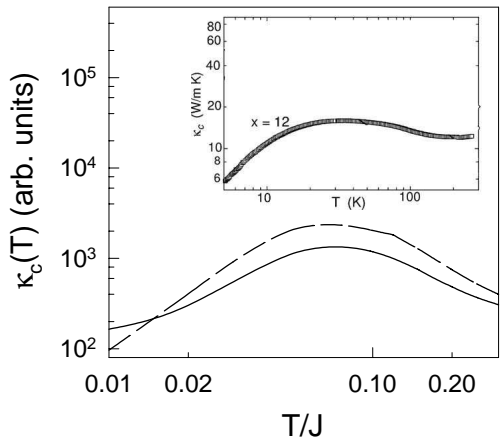}} \caption{The
thermal conductivity of the doped two-leg ladder material as a
function of temperature at doping $\delta=0.16$ (solid line) and
$\delta=0.20$ (dashed line) for parameter $t/J=2.5$. Inset: the
experimental result on Sr$_{14-x}$Ca$_{x}$Cu$_{24}$O$_{41}$ with
$x=12$ taken from Ref. \cite{sologubenko}.}
\end{figure}

In the two-leg ladder material
Sr$_{14-x}$Ca$_{x}$Cu$_{24}$O$_{41}$, the most remarkable
expression of the nonconventional physics is found in the doped
regime where the hole doping concentration on ladders at $x=8\sim
12$ is corresponding to $\delta=0.16\sim 0.20$ per Cu ladder
\cite{dagotto,uehara,nagata}. In this doped regime,
superconductivity appears under pressure in low temperatures, and
the value of $J_{\parallel}$ has been estimated \cite{katano} as
$J=J_{\parallel}\approx 90$ meV $\approx 1000$K. Therefore in the
following discussions, we focus on this doped regime. In Fig. 1,
we present the results of the thermal conductivity
$\kappa_{c}(\omega)$ in Eq. (6a) as a function of frequency at
$\delta=0.16$ (solid line) and $\delta=0.20$ (dashed line) for
$t/J=2.5$ with $T=0.05J$. Although $\kappa_{c}(\omega)$ is not
observable from experiments, its features will have observable
implications on the observable $\kappa_{c}(T)$. Our results in
Fig. 1 show that the frequency dependence of the thermal
conductivity spectrum is characterized by a rather sharp
low-energy peak. The position of this low-energy peak is doping
dependent, and is located at a finite energy $\omega\sim
0.1t=0.25J\approx 23$ meV. This low-energy peak is corresponding
to the peak observed in the conductivity spectrum
\cite{osafune,feng2}. Moreover, we also find from the above
calculations that although both dressed holons and spinons are
responsible for the thermal conductivity $\kappa_{c}(\omega)$, the
contribution from the dressed spinons is much larger than that
from the dressed holons, i.e.,
$\kappa^{(s)}_{c}(\omega)\gg\kappa^{(h)}_{c} (\omega)$, and
therefore the thermal conductivity of the doped two-leg ladder
antiferromagnet in the hole doped regime $\delta=0.16\sim 0.20$ is
mainly determined by its dressed spinon part
$\kappa^{(s)}_{c}(\omega)$.

Now we turn to discuss the temperature dependence of the thermal
conductivity $\kappa_{c}(T)$, which is observable from experiments
and can be obtained from Eq. (6a) as $\kappa_{c}(T)=
\lim_{\omega\rightarrow 0}\kappa_{c}(\omega,T)$. The results of
$\kappa_{c}(T)$ at $\delta=0.16$ (solid line) and $\delta=0.20$
(dashed line) for $t/J=2.5$ are plotted in Fig. 2 in comparison
with the corresponding experimental results \cite{sologubenko}
taken on Sr$_{14-x}$Ca$_{x}$Cu$_{24}$O$_{41}$ (inset), where the
hole doping concentration on ladders at $x=12$ is $\delta=0.20$
per Cu ladder. The present results indicate that the temperature
dependence of the thermal conductivity in the normal-state
exhibits a broad band, i.e., $\kappa_{c}(T)$ increases
monotonously with increasing temperature at low temperatures
$T<0.05J\approx 50$K, and is weakly temperature dependent in the
temperature range 0.05$J\approx 50$K$<T< $0.1$J \approx 100$K,
then decreases for temperatures $T>0.1J \approx 100$K, in
qualitative agreement with the experimental data
\cite{sologubenko}. For $T\leq 0.01J \approx 10$K, the system is a
superconductor under pressure. In this case, the thermal
conductivity of the doped two-leg ladder antiferromagnet is under
investigation now.

In the above discussions, the central concern of the heat
transport in the doped two-leg ladder material
Sr$_{14-x}$Ca$_{x}$Cu$_{24}$O$_{41}$ is the charge-spin
separation, then the contribution from the dressed spinons
dominates the thermal conductivity. Since
$\kappa^{(s)}_{c}(\omega,T)$ in Eq. (6c) is obtained in terms of
the dressed spinon longitudinal and transverse Green's functions
$D_{L}(k,\omega)$ and $D_{T}(k,\omega)$, while these dressed
spinon Green's functions are evaluated by considering the
second-order correction due to the dressed holon pair bubble, then
the observed unusual frequency and temperature dependence of the
thermal conductivity spectrum of the doped two-leg ladder
antiferromagnet is closely related to the commensurate spin
response \cite{nagata,feng3}. The dynamical spin structure factor
of the doped two-leg ladder antiferromagnet has been obtained
within the charge-spin separation fermion-spin theory \cite{feng3}
in terms of the full dressed spinon longitudinal and transverse
Green's functions as,
\begin{mathletters}
\begin{eqnarray}
S(k,\omega)&=&-2[1+n_{B}(\omega)][2{\rm Im}D_{L}(k,\omega)+2{\rm
Im}D_{T}(k,\omega)]\nonumber\\
&=& [1+n_{B}(\omega)][A_{L}^{(s)} (k,\omega)+
A_{T}^{(s)}(k,\omega)]\nonumber \\
&=&-{4[1+n_{B}(\omega)](B^{(1)}_{k})^{2}{\rm Im} \Sigma_{LT}^{(s)}
(k,\omega)\over[W( k,\omega)]^{2}+[B^{(1)}_{k}{\rm
Im}\Sigma_{LT}^{(s)}(k, \omega)]^{2}},\\
W( k,\omega)&=&\omega^{2}-(\omega^{(1)}_{k})^{2}-B^{(1)}_{k}{\rm
Re}\Sigma_{LT}^{(s)}(k, \omega),
\end{eqnarray}
\end{mathletters}
where ${\rm Im}\Sigma_{LT}^{(s)}(k,\omega)={\rm Im}
\Sigma_{L}^{(s)}(k,\omega)+{\rm Im}\Sigma_{T}^{(s)}(k,\omega)$,
${\rm Re}\Sigma_{LT}^{(s)}(k,\omega)={\rm Re}\Sigma_{L}^{(s)}
(k,\omega)+{\rm Re}\Sigma_{T}^{(s)}(k,\omega)$, while ${\rm
Im}\Sigma^{(s)}_{L}(k,\omega)$ (${\rm Im}\Sigma^{(s)}_{T}(k,
\omega)$) and ${\rm Re} \Sigma^{(s)}_{L}(k,\omega)$ (${\rm
Re}\Sigma^{(s)}_{T}(k,\omega)$) are the imaginary and real parts
of the second order longitudinal (transverse) spinon self-energy
in Eqs. (8c) and (8d). The renormalized spin excitation
$E^{2}_{k}= (\omega^{(1)}_{k})^{2}+ B^{(1)}_{k}{\rm
Re}\Sigma^{(s)}_{LT}(k,E_{k})$ in $S(k,\omega)$ is doping, energy,
and interchain coupling dependent. In the two-leg ladder system,
the quantum interference effect between the chains manifests
itself by the interchain coupling, i.e., the quantum interference
increases with increasing the strength of the interchain coupling.
In the materials of interest, $J_{\perp}\sim J_{\parallel}$ and
$t_{\perp}\sim t_{\parallel}$, we have shown in detail in Ref.
\cite{feng3}, the dynamical spin structure factor in Eq. (12) has
a well-defined resonance character, where $S(k,\omega)$ exhibits
the commensurate peak when the incoming neutron energy $\omega$ is
equal to the renormalized spin excitation $E_{k}$, i.e., $W(
k_{c},\omega)\equiv [\omega^{2}- (\omega^{(1)}_{k_{c}})^{2}
-B^{(1)}_{k_{c}}{\rm Re} \Sigma^{(s)}_{LT}(k_{c},\omega)]^{2} =
(\omega^{2}-E^{2}_{k_{c}})^{2}\sim 0$ at antiferromagnetic wave
vectors ${\bf k}_{{\rm AF}}$, then the height of this commensurate
peak is determined by the imaginary part of the spinon self-energy
$1/{\rm Im} \Sigma^{(s)}_{LT}(k_{c},\omega)$. Since the incoming
neutron resonance energy $\omega_{{\rm r}} =E_{k}$ is finite
\cite{feng3}, then there is a spin gap in the system. This spin
gap leads to that the low-energy peak in $\kappa_{c}(\omega)$ is
located at a finite energy $\omega_{{\rm peak}}\approx 23$meV.
This anticipated spin gap $\Delta_{S}\approx 23$ meV is not too
far from the spin gap $\approx 32$ meV observed \cite{katano} in
Sr$_{14-x}$Ca$_{x}$Cu$_{24}$O$_{41}$. Therefore the commensurate
spin fluctuation of the doped two-leg ladder material
Sr$_{14-x}$Ca$_{x}$Cu$_{24}$O$_{41}$ is responsible for the heat
transport, in other words, the energy transport via magnetic
excitations dominates the thermal conductivity \cite{sologubenko}.
Based on the simple theoretical estimate \cite{hess}, the large
thermal conductivity observed in the two-leg ladder material
Sr$_{14}$Cu$_{24}$O$_{41}$ has been interpreted in terms of the
magnetic excitations in the system \cite{hess}. Our present
explanation is also consistent with theirs \cite{hess}.

To conclude we have discussed the heat transport of the two-leg
ladder material Sr$_{14-x}$Ca$_{x}$Cu$_{24}$O$_{41}$ within the
$t$-$J$ ladder in the doped regime where superconductivity appears
under pressure in low temperatures. It is shown that the energy
dependence of the thermal conductivity spectrum
$\kappa_{c}(\omega)$ shows a low-energy peak, and the position of
this low-energy peak is doping dependent, and is located at a
finite energy. The temperature dependence of the thermal
conductivity $\kappa_{c}(T)$ is characterized by a broad band.
$\kappa_{c}(T)$ increases monotonously with increasing temperature
at low temperatures $T<0.05J$, and is weakly temperature dependent
in the temperature range $0.05J<T<0.1J$, then decreases for
temperatures $T>0.1J$. Our result of the temperature dependence of
the thermal conductivity is in qualitative agreement with the
major experimental observations of
Sr$_{14-x}$Ca$_{x}$Cu$_{24}$O$_{41}$ in the normal-state
\cite{sologubenko}. Our result also shows that although both
dressed holons and spinons are responsible for the thermal
conductivity, the contribution from the dressed spinons dominates
the heat transport.

\acknowledgments The authors would like to thank Dr. Ying Liang,
Dr. Tianxing Ma, and Dr. Yun Song for the helpful discussions.
This work was supported by the National Natural Science Foundation
of China under Grant Nos. 10125415 and 90403005, the Grant from
Beijing Normal University, and the National Science Council.

\end{document}